\def\breakon{\end{multicols}\widetext\vspace{-.2cm}
\noindent\rule{.48\linewidth}{.3mm}\rule{.3mm}{.3cm}\vspace{.0cm}}
\def\breakoff{\vspace{-.2cm}
\noindent
\rule{.52\linewidth}{.0mm}\rule[-.27cm]{.3mm}{.3cm}\rule{.48\linewidth}{.3mm}
\vspace{-.3cm}
\begin{multicols}{2}
\narrowtext}
\documentstyle[aps,prl,multicol]{revtex}

\begin{document}

\draft

\widetext

\title{Interaction effects on quasiparticle localization in dirty
  superconductors}

\author{M. Jeng$^1$, A. W. W. Ludwig$^1$, T. Senthil$^{2}$, C. Chamon$^3$}

\address{
$^1$ Dept. of Physics, University of California, Santa Barbara, CA 93106\\
$^2$ Dept. of Physics, Massachusetts Institute of Technology, \\
77 Massachusetts Ave., Cambridge, MA 02139\\
$^3$ Dept. of Physics, Boston University, Boston, MA 02215}

\maketitle

%%%%%%%%%%%%%%%%%%%%%%%%%%%%%%%%%%%%%%%%%%%%%%%%%%%%%%%%%%%%%%%%%%%%%%%%%%%%%%
\begin{abstract}
  We study how quasiparticle interactions affect their localization
  properties in dirty superconductors with broken time reversal 
  symmetry -- for
  example in a magnetic field.  For $SU(2)$ spin-rotation invariant (class C)
  systems, the only important coupling is the spin-spin triplet interaction,
  which we study within a renormalization group approach. Either an
  additional Zeeman coupling or a complete breaking of spin rotation symmetry
  renders all interactions irrelevant. These two situations realize,
  respectively, the non-interacting unitary Anderson and the ``thermal'' (class
  D) universality class. Our results imply a stable metallic phase in 2D for
  class D. Experimental implications are discussed.
\end{abstract}

\pacs{PACS: 71.30.+h, 71.23.An}

% 71.10.-w Theories and models of many electron systems
% 71.23.-k Electronic structure of disordered solids
% 71.30.+h Metal-insulator transitions and other electronic transitions
% 71.23.An Theories and models; localized states
% 73.23.-b Mesoscopic systems

%%%%%%%%%%%%%%%%%%%%%%%%%%%%%%%%%%%%%%%%%%%%%%%%%%%%%%%%%%%%%%%%%%%%%%%%%%%%%%

\begin{multicols}{2}

\narrowtext

%%%%%%%%%%%%%%%%%%%%%%%%%%%%%%%%%%%%%%%%
% 
%%%%%%%%%%%%%%%%%%%%%%%%%%%%%%%%%%%%%%%%

The interplay between disorder and interactions in electronic systems
underlies several interesting phenomena in solids. A famous, though poorly
understood, example is the metal insulator transition in three-dimensional
dirty solids. Although much theoretical progress has been achieved in
understanding the Anderson localization transition of {\it non-interacting}
electrons, direct contact with experiments has been problematic due to the
complicating effect of interactions. Interesting field theoretic attempts
have been made to incorporate interactions into the successful scaling theory
of the Anderson transition \cite{Belitz}. However the resulting theories are
complicated to analyse, and in most cases have thus far not led to a good
description of real metal-insulator transitions. In two dimensional systems
(2D), the situation is even worse: even the possible stability of a genuine
metallic phase in the presence of disorder and interactions is a matter of
considerable current debate \cite{review}.  
Recent work\cite{BK1,CM} has made
some progress in understanding the field theory of the disordered,
interacting, two-dimensional electron gas.

As recently emphasized, the dynamics of quasiparticles in a superconductor
provides a new, though still experimentally relevant, context to address
localization issues
\cite{Zirnbauer,SFBN,noSpin,Smitha,Bocquet,Donecomponent}.  All disordered
superconductors fall into one of two categories according to the nature of
their quasiparticle transport properties -- superconducting ``thermal metals''
(with delocalized quasiparticles) or ``thermal insulators'' (with localized
quasiparticles). Interesting differences arise with localization physics in a
normal metal due to the lack of conservation of the quasiparticle electric
charge in the superconductor.

In this paper, we take up the task of describing quasiparticle localization
inside superconductors in the presence of both disorder and interactions
\cite{Bull}. We argue that this problem is simpler than the corresponding
problem in normal metals. We explicitly identify physical situations in which
the interaction effects are unimportant for the long distance physics. The
corresponding experimental systems thus provide a clear opportunity to study
Anderson localization transitions, unhindered by interaction effects.  We
also demonstrate the stability of a thermal metal phase in 2D
in the presence of both interactions and disorder inside
superconductors under appropriate conditions. The insights
gained from studying the superconductor may be valuable in developing an
understanding of the normal metal.
 
Within the standard mean field treatment of pairing, the dynamics of
non-interacting BCS quasiparticles is governed by a quadratic
Bogoliubov-deGennes (BdG) Hamiltonian, subject to static disorder in the
normal and  pairing potentials. The localization physics 
of quasiparticles has been studied previously in this approximation. 
A total of four universality
classes, different from the three known standard classes for normal metals,
have been found for the possible localization behavior of 
non-interacting quasiparticles inside dirty 
 superconductors\cite{Zirnbauer,SFBN,noSpin}, 
on length scales much larger than the mean free path. 
These correspond to BdG Hamiltonians with or without
spin-rotation and/or time reversal symmetries.
Since charge is not conserved in a superconductor, the nature
of a phase (metal, insulator, or Hall insulator) manifests itself not in
charge transport, but rather in thermal-transport or, when spin is conserved,
in concomitant spin-transport.

Here we investigate the effect of quasiparticle 
interactions, focusing on situations lacking time reversal invariance.
One simplication offered by the 
superconductor (as compared to a normal metal) is that the long-range
Coulomb interaction is always screened out by the condensate. Thus, the
quasiparticle interaction is short-ranged. Moreover, lack of charge
conservation renders the singlet density-density interaction unimportant
altogether. As in normal metals, systems with broken time reversal symmetry
offer the further simplification that interactions in the Cooper channel 
are also unimportant. The most significant interaction then is a 
short-ranged interaction between the quasiparticle spin densities. 

Possible experimental realizations include certain heavy fermion
superconductors\cite{HeavyFermion}, which typically have strong spin-orbit
(S.O.) scattering, and superfluid He-3 in porous media\cite{Helium3}; these
fall into class D of \cite{Zirnbauer}.  

Perhaps the most promising prospect
is Type II superconductors in strong magnetic fields~\cite{Smitha}. It has
been suggested that a ``thermal insulator-metal'' transition, which can be
probed by ultralow temperature heat transport measurements, may be driven in
such a superconductor by simply changing the magnetic field.  
If the Zeeman coupling to the magnetic field and S.O. scattering can be
ignored, the transition is in the universality class describing a
superconductor with full $SU(2)$ spin symmetry, but  without 
time reversal symmetry
(class C of Ref.\cite{Zirnbauer}).  However, typical Zeeman energies are
expected to be much bigger than the low temperatures necessary to clearly
extract the electronic heat transport. Thus, the asymptotic critical
properties will be described by a theory that includes the Zeeman
coupling. In the absence of S.O.  scattering this situation is formally in
the same universality class as that of spinless {\it electrons} in a magnetic
field\cite{SFBN}. Short-range interactions are known to be irrelevant in the
latter model\cite{ShortRange} - a physical consequence of the Pauli
principle. 
Therefore, the field-driven thermal metal-insulator transition in a bulk 3D
type II superconductor provides an excellent 
(and possibly unique) opportunity to experimentally study the
 non-interacting 3D (unitary) 
Anderson transition, in contrast to normal metals where
the long-range Coulomb interaction changes the universality class. Despite
being irrelevant, the short-range interactions in this system can still
affect the low frequency, finite temperature ($T$) dynamics: 
 a calculation along the lines of \cite{wang} shows that
the thermal conductance $\kappa$ for $T\to 0$ behaves as $\kappa/T\propto
T^{\theta (d-2)}$. Without interactions,  $\theta=\nu$, the localization length exponent.
With interactions, $\theta= min(\nu,\frac{p}{2})$,
where $p$ is the ``dephasing'' exponent arising from the irrelevant
interactions. To lowest order in the $d=2+\epsilon$ expansion one finds  $p<2\nu$
($p=1.3$ when naively extrapolated to $d=3$.)

If, on the other hand, even uniaxial
spin rotation symmetry is broken (class D of \cite{Zirnbauer}) 
due to, for example, S.O
scattering, then for the same physical reason as above, interactions are
again expected to be  irrelevant. 
In 3D, this
is thus expected to realize the thermal metal-insulator transition discussed
in \cite{noSpin}.  In 2D, this implies that the (thermal) metallic phase
in non-interacting models of such
superconductors\cite{noSpin,Bocquet,Donecomponent} is stable to the inclusion
of interactions. This then provides a concrete theoretical instance
of a stable metallic phase in 2D, albeit for thermal transport.

To study interaction effects in classes C and D, we construct non-linear
sigma models[NL$\sigma$M] which generalize those constructed by Finkelstein for normal
metals\cite{Belitz}.  Within this framework  we see the 
irrelevance of all interactions in class D, and that of singlet and Cooper
interactions in class C.  For the latter case we carry out a perturbative
Wilsonian renormalization group (RG) analysis:  In $2+\epsilon$ dimensions
($\epsilon>0$) the remaining triplet interaction is found to be marginal, to
1-loop order, at the fixed point describing the (thermal) metal-insulator
transition without interactions\cite{Smitha}.  In 2D, we find that the
triplet interaction strongly affects the weak localization correction to the
spin- (and thermal-) conductivity in the metallic phase (at weak coupling),
which changes sign for sufficiently attractive interactions.

%%%%%%%%%%%%%%%%%%%%%%%%%%%%%%%%%%

{\bf Class C:} 
We start with a general BCS Hamiltonian for a dirty singlet superconductor,
possessing spin rotation invariance, but no time reversal symmetry, in the
{\it absence}  {\it of quasiparticle interactions}: \breakon
\begin{equation}
\label{HCclass}
{\mathcal H} = \int d^d x \sum_{\alpha} \
   \psi_\alpha^{\dagger} 
   \left(-\frac{\nabla^2}{2m}-\mu+V(x)\right)\psi_\alpha +
   \left(\Delta (x)\psi_{\uparrow}^{\dagger}\psi_{\downarrow}^{\dagger}
   + \mathrm{h.c.}\right)
\end{equation}
Here $\alpha= \uparrow,\downarrow $ labels the spin. The potential
$V(x)$, as well as the real and imaginary parts of the complex gap function
$\Delta (x)$, are (static) random variables with Gaussian distributions of
zero mean and (without loss of
generality) of same variance.  The vanishing mean  of the complex $\Delta(x)$ is physically
appropriate in the Type II superconductor in the mixed phase with randomly
located vortices, and in other physical situations. We treat the disorder
average using the Schwinger-Keldysh method (described for the unitary
symmetry class 
\cite{CLN} - see also \cite{Kamenev}). 
Within this formalism, the fermions acquire an
additional Keldysh index $i= 1,2$ denoting the time-ordered and
anti-time-ordered branches of the path integral.  At zero temperature the
Keldysh functional integral action corresponding to the above Hamiltonian is
(spin and Keldysh labels supressed where possible)
\begin{eqnarray}
\nonumber
S & = &\int d^dx dt \  \psi^{\dagger} (x,t) \sigma_z 
        \left[i\partial_t-\frac{\nabla^2}{2m} - \mu\right] \psi (x,t) 
 + i\eta \int  d^d x \frac{d\omega}{2\pi}
         \ \mathrm{sgn} (\omega) \psi^{\dagger}(x,\omega) \psi(x,\omega) + \\ 
 & & + \int d^d x dt \left[
        V(x) \psi^{\dagger} (x,t) \sigma_z \psi (x,t) +
        \Delta (x)\psi^{\dagger}_{\uparrow} (x,t) \sigma_z 
                \psi^{\dagger}_{\downarrow} (x,t) +
        \Delta^{*}(x) \psi^{\dagger}_{\downarrow} (x,t) \sigma_z 
                \psi^{\dagger}_{\uparrow} (x,t) \right] 
\end{eqnarray}

\breakoff

\noindent
where $\sigma_z$ is a Pauli matrix in Keldysh space.
The second  term ($\eta>0$) specifies the
time-ordering on the Keldysh contour. 
The symmetries are made explicit by introducing
a 4-component field on each Keldysh branch,
\begin{equation}
\label{eightcomponent}
\chi_i(x,t) 
=
\pmatrix{
\chi_{i, a=1}(x,t)
\cr
\chi_{i, a=2}(x,t)
\cr} \equiv 
 {1\over \sqrt{2}} \pmatrix{
\psi_i(x,t) \cr
i \tau_y 
\psi_i^\dagger(x,t) \cr}
\end{equation}
satisfying the reality condition $\chi_i^\dagger =(C \chi_i)^T$ with $C=\mu_y
\tau_y$ (Pauli matrices ${\vec \mu}_{ab}$ act on the `particle-hole' index
$a$ of $\chi_i$, and ${\vec \tau}_{\alpha \beta}$ on the spin-index.)
  When expressed in terms of the 8-component fermion field
 $\chi(x,\omega)$, the Keldysh action above is readily seen to be invariant
 under independent symplectic transformations $U(\omega)\in Sp(4) \subset
 U(4) $ for each frequency $\omega$ (repeated indices summed),
$$
\chi_{i,a,\alpha}(x,\omega)\to  \  U_{i\alpha; j \beta}(\omega)
\ \  \chi_{j,a,\beta}(x,\omega); \ \ 
U^T \tau_y\sigma_z U = \tau_y\sigma_z
$$
when $\eta=0$. A finite $\eta$ breaks the symmetry down to a $U(2)$
subgroup.  By standard arguments the low energy degrees of freedom are
described by a diffusion mode matrix field $Q(x) =Q^{ij;
\alpha,\beta}_{\omega_1,\omega_2}(x)$ which decouples fermion 
bilinears
$  \sigma_z^{ij}
\bigl (\chi^\dagger_{i a \alpha}(x,\omega_1)
\chi_{j a \beta}(x,\omega_2) \big ) $,
appearing after  the disorder average.
It carries
 spin ($\alpha,\beta$), Keldysh
($i,j$), and frequency indices.
$Q$ takes values on the saddle point
manifold specified in Eq.~(\ref{Qconstraints}) below, and its dynamics is
governed by the Keldysh NL$\sigma$M  action Eq.~(\ref{SKeldysh}).
This formulation of the non-interacting disordered theory is
entirely equivalent to those obtained using the replica trick or supersymmetry.

{\it Effect of interactions:}
In the absence of disorder, the interactions at the Fermi surface
can be, most generally\cite{ShankarRG}, of
density-density (singlet), spin-spin (triplet) and (singlet and triplet)
Cooper  types.  
Introducing three Hubbard-Stratonovich (H.S.)
fields to decouple these  four-fermion interactions,
and condensing the H.S. field which
decouples the Cooper interactions (gap function)
in the spin-singlet channel, 
gives the BCS mean field theory.
In the presence of disorder, but no 
interactions,  this has the form of Eq.(\ref{HCclass}).
Quasiparticle interactions arise from the (dynamical) fluctuations
of the three H.S. fields.
In particular, 
residual interactions  in the Cooper channel
arise from the fluctuations of the amplitude  and the phase of
the gap function.  The gapless phase fluctuations
can be seen to decouple from the BdG quasiparticles at asymptotically low
energies, and their only effect is to render a long-range
Coulomb interaction short-ranged\cite{longpaper}.
This confirms the  expected screening by the condensate, mentioned above.
Averaging over disorder in the presence of the three  H.S. fields yields, 
along the lines of\cite{CLN},
 a theory of diffusion modes $Q$ interacting with the latter.
In the absence of 
fluctuations of the gap function amplitude, the remaining two
H.S. fields can be integrated out straightforwardly. The singlet
H.S. field  is seen to decouple, and integration over the
triplet  H.S. field leads to an interaction term of
the familiar Finkelstein form. In particular, the resulting
Keldysh-Finkelstein action  reads (after simple rescalings and unitary transformations):
\begin{equation}
\label{ZKeldysh1}
Z   =   \int [{\mathcal{D}} Q] e^{-S[Q]}; \qquad \qquad S[Q]  =  S_D [Q] + S_{\mathrm{int}}[Q] \\
\end{equation}
\begin{equation}
\label{SKeldysh}
S_D [Q]  =  
 \int { d^dx\over 4} 
     \left[ {1\over 8 \pi g} \  Tr (\nabla Q)^2 + 4z Tr(i\omega\sigma_z-
        \eta\mathrm{sign}(\omega))Q\right] 
\end{equation}
\begin{equation}
\label{Finkelsteinterm}
S_{int}[Q]  =  i \pi U_t z^2 \sum_i \sigma_z^{ii} \int d^d x \int  dt \ 
Q^{ii, \alpha \beta}_{t, t}
Q^{ii, \beta \alpha}_{t, t}
\end{equation}
Here $Tr$ denotes the trace over all indices of $Q$,
including the frequency index $\omega$, and
$ Q^{ii, \alpha \beta}_{t_1, t_2}$  is the corresponding (double-)
 time Fourier transform.
  The saddle point   manifold of massless modes of the NL$\sigma$M
is described in frequency space by
\begin{eqnarray}
\label{Qconstraints}
Q^2=1, \;\; Q^{\dagger}=Q, \;\; Tr(Q)=0; \  
\;\; \tau_y \Sigma_x Q \tau_y \Sigma_x = -Q^{T}
\end{eqnarray}
\noindent where $\Sigma_x$  exchanges
positive and negative frequencies.
Note that $ Q^{ii, \alpha \beta}_{t, t}$  
transforms as a spin-triplet. Consequently, 
 Eq.(\ref{Finkelsteinterm}) is the
only  $SU(2)$ spin-rotation invariant Finkelstein interaction term that can be written.
This was expected: due to the lack of charge conservation there is no
massless charge diffusion mode, and due to the lack of time-reversal
symmetry, there is no massless Cooperon mode on the saddle point manifold.
Hence Eq.(\ref{Finkelsteinterm})  represents the only non-vanishing
interaction, which is that  between the spin diffusion modes.
Here $1/g$ is proportional to the  (spin) conductivity, and $U_t (<0)$ is the
(repulsive) triplet interaction strength.

We perform a perturbative (Wilsonian) R.G. analysis,
parametrizing the saddle point manifold by
independent matrix fields $V$,  similar to \cite{CLN}.
The  wavefunction renormalization  is found by evaluating
$\langle Q\rangle$, using the same logic as
in the standard O(N) NL$\sigma$M~\cite{Amit}.
The 1-loop  R.G.  equations, valid 
to lowest order in $g$ (and $z$), but to all orders in $U_t z$,
are found in $d = 2+\epsilon$: 
\begin{eqnarray}
&&
{d g\over dl}=-\frac{\epsilon}{2} g + 
        \left[7-6\left(1-\frac{1}{2U_tz}\right)\log (1-2U_tz)\right]g^2 +O(g^3)
\nonumber  \\
&&
{d (U_t z)\over dl}
 = O(g^2), \qquad 
{d z \over dl} =\left[\frac{\epsilon}{2} - (1 + 6U_t z) g  \right] z  + O(g^2) \nonumber
\end{eqnarray}
 ($l$ is twice the log of the length scale). These are
extracted from  the renormalization of the terms quadratic in $V$
in the action.
We also confirmed that the same R.G. equations are obtained from the renormalization of
the terms of next higher order in $V$. This is a necessary condition for the
renormalizability of the theory. 

Note that in contrast to other universality classes, 
$U_t z$, a measure of the interaction strength, does not renormalize to 1-loop
order in class C.  

Without interactions ($U_t=0$), we recover the non-interacting
fixed point at 
$g=\frac{\epsilon}{2}+...$,
 which is the thermal (or spin) metal-insulator
transition considered in~\cite{Smitha}. We see from the second R.G.
 equation that  the interaction $U_t z $ is marginal at this fixed point, 
at least to 1-loop order. 
Hence, showing that interactions are relevant or irrelevant would require
working to higher-loop order. 

Next we specialize to 2D, focussing on the metallic phase where
$g\ll1$.  Since $U_t z$ does not renormalize to lowest order in $g$, it will
be constant over a wide range of length scales.  Hence, in this range, the
thermal- ($\kappa$), and spin- ($\sigma^s$) conductivity is given by the
1-loop result (${3 \hbar^2 \over 4 \pi^2 k_B^2} \ { \kappa \over
  T}=\sigma^s$):
$$
\sigma^s =
 \sigma^s_0 - \frac{1}{4 \pi^2} 
        \left[ 7-6\left(1-\frac{1}{2U_tz}\right)\log (1-2U_tz) \right]
        \log \left( \frac{L}{\ell_e} \right) 
$$
Here $\sigma^s_0$ is the (bare) spin conductivity 
at the scale of the elastic mean-free path
 $\ell_e$, and
$\sigma^s$ its value at length scale $L$. 
This expression  is perturbative in $g=({ 2\over \pi}) 1/ \sigma^s$, 
 but $U_t z$  need not be small.
(Note that the quantity in square brackets goes to 1 as interactions
are removed.)
Thus, a  repulsive (attractive)  triplet
interaction is seen 
 to decrease (enhance) the weak localization
correction to the  (spin-, or thermal-) conductivity,
as compared to the non-interacting case.
For sufficiently strongly repulsive interactions, $2U_t z <
-0.37..$,  the  $g^2$-coefficient in $dg/dl$ changes sign, 
implying that   $g$ flows back to weak coupling, 
at least until, potentially, higher-loop effects set in which
may reverse the flow.

{\bf Class D:}  Turning now to  situations
lacking spin-rotation as well as  time
reversal symmetry, we 
can see, without extensive calculations, that
all Finkelstein-type interaction terms  are   absent.
In this case, the  diffusion mode matrix field 
$Q^{ij}_{t_1, t_2}(x)$ decoupling fermion 
bilinears
$  \sigma_z^{ij}
\bigl (\chi^\dagger_{i a \alpha}(x,t_1)
\chi_{j a \alpha}(x,t_2) \big ) $
carries only Keldysh ($i,j$) and frequency, but no  spin indices.
Repeating the steps above Eq.(\ref{ZKeldysh1}),  both the singlet
and triplet H.S. fields are seen to decouple. This
was expected since now, due to lack of spin-rotational symmetry,
also the spin diffusion mode is absent from the saddle point manifold
of massless modes.
Indeed, the only possible Finkelstein interaction term as in Eq.(\ref{Finkelsteinterm})
(but without spin indices, $\alpha, \beta$) vanishes due to the
antisymmetry of $Q^{ii}_{t_1, t_2}(x)$, which follows from the reality
condition below Eq.(\ref{eightcomponent}).

Consequently, we conclude that
 the 3D thermal metal-insulator transition in class
D\cite{noSpin} is
expected to be  unmodified by quasiparticle interactions, as mentioned above. 
In 2D, in the absence of interactions, a (thermal) metallic phase,
stable to quantum interference, 
can exist inside generic superconductors lacking spin rotation and time
reversal invariances\cite{noSpin,Bocquet}.  
 (For spin-polarized versions, see Ref.\cite{Donecomponent}.)  
The absence of any type of marginal Finkelstein interaction terms in class D,
found above,  thus implies that this 2D  metallic
phase remains stable even upon inclusion of any type of interactions.

We end our discussion with a  general  caveat.
 The question as to whether or not additional potentially `dangerous', 
 i.e. relevant or marginal, 
long-range (in time)  versions  of the interaction terms
are eventually generated upon the R.G.,  hinges upon a proper
understanding of the renormalizability of the whole class of
 Finkelstein-type theories in general, which is currently lacking.
 However, it is  not expected that initially short-range 
interactions would generate  such long-range ones, upon integration over short-distance,
short-time fluctuations in an R.G. transformation.
These issues will be addressed in more detail in Ref.\cite{longpaper}.

After our work was completed, and some of our results were reported
in\cite{Bull}, we learned about the work by Fabrizio {\it et
  al.} \cite{Fabrizio}, whose results for the correction to the 2D
conductivity and density of states for class C are in agreement with ours.

We thank Patrick Lee, Christopher Mudry and Chetan Nayak  for useful discussions.
This work was supported by NSF
grants No.  DMR-00-75064 (A.W.W.L, M.J), and DMR-98-76208 (C.C.). The work of
T.S. was supported by the MRSEC program of the NSF through the grant
DMR-98-08941. C.C. acknowledges the support from the Alfred P. Sloan
Foundation.

%%%%%%%%%%%%%%%%%%%%%%%%%%%%%%%%%%

\end{multicols} 

\end{document}